\documentclass[twocolumn,fleqn,,usenatbib]{mnras}
\usepackage{hyperref}
\usepackage{bm}
\usepackage{graphicx}
\usepackage{multirow,tabularx}
\usepackage{enumitem}
\usepackage[flushleft]{threeparttable}
\usepackage[]{color}

\usepackage{natbib}
\usepackage{float}
\usepackage{breqn}
\usepackage[T1]{fontenc}
\usepackage[utf8]{inputenc}
\usepackage[table, dvipsnames]{xcolor}

\usepackage{amsfonts,amsmath,amssymb,esint}

\usepackage{color}
\usepackage{xcolor}
\usepackage{url}

\title[X-ray Irradiated AGN Winds]{Time-Dependent AGN Disc Winds II - Effects of Photoionization}

\author[S. Dyda et al.]{
Sergei Dyda,$^{1,2,3}$\thanks{sdyda@ua.edu}
Randall C. Dannen,$^{3}$
Timothy R. Kallman,$^{4}$
Shane W. Davis,$^{1,2}$
and Daniel Proga$^{3}$
\\
$^{1}$ Department of Astronomy, University of Virginia, 530 McCormick Rd., Charlottesville, VA 22904, USA \\
$^{2}$ Virginia Institute for Theoretical Astronomy, University of Virginia, Charlottesville, VA 22904, USA \\
$^{3}$ Department of Physics \& Astronomy, University of Alabama, Gallalee Hall, 514 University Blvd, Tuscaloosa, AL 35401, USA \\
$^{4}$ Department of Physics \& Astronomy, University of Nevada, Las Vegas, 4505 S. Maryland Pkwy, Las Vegas, NV, 89154-4002, USA \\
$^{5}$ NASA Goddard Space Flight Center, Greenbelt, MD 20771, USA
}

\begin{document}

\label{firstpage}
\pagerange{\pageref{firstpage}--\pageref{lastpage}}

\maketitle
\begin{abstract}
We use a combination of radiation hydrodynamics (rad-HD) and photoionization modeling to study line-driven disc winds for a range of black hole masses. 
We refined previous models by incorporating heating, cooling, and radiation forces from spectral lines calculated using a photoionization code, assuming that composite AGN spectra irradiate the gas. 
For black holes with masses  $3 \times 10^{6} \lesssim {\rm M_{BH}/M_{\odot}} \lesssim 10^{8}$, the mass loss rate, ${\rm \dot{M}_w}$ increases proportionally with the disk Eddington fraction, $\Gamma$. The insensitivity of ${\rm \dot{M}_w}$ to the hardness of the spectral energy distribution (SED) arises because the central region is dominated by radiation in the frequency range 
with ample spectral lines for the range of $M_{BH}$ considered here.
Disc winds are suppressed or fail outside the above mass range because of a dearth of line-driving photons.  
We find \emph{stronger} winds, both in terms of ${\rm \dot{M}_w}$ and wind velocity compared to previous disc wind models. 
Our winds are stronger because of an enhanced line force from including many spectral lines in the X-ray band. These lines were unavailable and, hence, unaccounted for in previous photoionization studies and their subsequent application to AGN wind models. 
For $\Gamma \gtrsim 0.4$, ${\rm \dot{M}_w}$ is higher than the assumed disc accretion rate, implying that the wind feeds back strongly. Our findings indicate the necessity of utilizing comprehensive and current atomic data along with a more thorough approach to radiation transfer — both spatially and temporally — to accurately calculate the line force.
\end{abstract}

\section{Introduction}
\label{sec:introduction}

Active Galactic Nuclei (AGN) launch powerful outflows through a dynamic interplay of gravitational and electromagnetic forces associated with the supermassive black holes at their cores. As matter accretes onto the black hole, a portion of it forms a hot, dense accretion disk. The intense gravitational forces within this disk lead to the acceleration of charged particles, generating immense, $\sim$ few $ \times 10^7$ K temperatures and intense, $L \sim 10^{40-47}$ ergs, luminosity. This radiation spans the electromagnetic spectrum, and in particular, X-ray and ultraviolet (UV) radiation ionizes and heats the surrounding gas. The resulting ionized gas can be accelerated via radiation pressure on spectral lines, resulting in high-speed winds which propel material away from the nucleus. These outflows can reach velocities of tens of thousands of kilometers per second and are an important form of feedback for their host galaxies by allowing energy and material to be transported away from the innermost region near the central supermassive black hole (SMBH) to the interstellar medium (ISM) and even outside the host galaxy.


One possible mechanism for launching and accelerating powerful outflows is radiation pressure due to spectral lines. 
In this \emph{line driven wind} scenario, the radiation field couples to the gas in two important ways. Firstly, it causes changes in the ionization state of the gas and allows the absorption and emission of radiation for spectral lines due to many species of ions. Secondly, the radiation field can impart momentum on the gas via interactions through this plethora of spectral transitions, and this momentum transfer can be sufficient for the gas to launch and escape the AGNs gravitational potential. Carefully accounting for these two physical effects is challenging as both the ionization and momentum transfer are highly non-linear, receiving contributions from across the electromagnetic spectrum and dependent on the geometry of the irradiated flow.

There has been a great deal of effort in developing photoionization codes to determine the state of the gas (e.g., \textsc{XSTAR} \cite{XSTAR2001} and \textsc{CLOUDY} \cite{Cloudy23}).  For a given gas density, temperature, elemental abundances, and incident spectral energy distribution (SED), these codes compute the required ionic abundances and opacities to replicate spectral features seen in astrophysical objects. The key physics behind the models is based on experimental and theoretical transition probabilities between ionic states and their respective oscillator strengths, often colloquially referred to as the \emph{line list}. 

These photoionization calculations are often very costly and only get more so as more atomic data is included. Due to this fact, it is computationally prohibitive to run these photoionization codes in parallel with hydrodynamical modeling. To make these calculations more feasible, previous work has reduced and tabulated the large parameter space of photoionization models to incorporate them into hydrodynamics simulations. For instance, the state of the gas is characterized by its temperature $T$ and the effect of the SED is encapsulated in an ionization parameter
\begin{equation}
    \xi = \frac{\left( 4 \pi \right)^2 J_{\xi}}{n_H},
\end{equation}
where $J_{\xi}$ is the mean intensity of ionizing photons and $n_H$ is the number density of hydrogen nucleons. The upper bound for the wavelength corresponds to 13.6 eV, the ionization energy of hydrogen. Further, it is important to recognize that the photoionization calculation is highly non-linear, so approximating the high energy part of the SED using a single parameter $\xi$ is itself an approximation.

After performing a series of photoionization models, one can interpolate gas properties (heating rates, effective number of optically thick lines, i.e., the force multiplier, etc.) as a function of microscopic gas parameters, such as temperature and the ionization parameter. This was done in the seminal paper \cite{Stevens1990}, hereafter SK90, which fit their results for the force multiplier in terms of the ionization parameter. This parameterization of the force multiplier in terms of the ionization parameter served as the basis for many hydrodynamics models of AGN disc winds (\cite{PSK2000} hereafter PSK2000, \cite{PK04} hereafter PK04, \cite{DSP2024} hereafter Paper I). Such models used ever more sophisticated means of estimating the ionization parameter via modeling the local ionizing flux. For instance, PK04 assumed the ionizing rays traversed along radial rays from the central source and were attenuated by the intervening wind via an electron scattering opacity. In Paper I, we solved the full time-dependent radiation transfer equation, including effects for scattering and absorption opacities. Despite these efforts, the models are ultimately limited by the underlying assumptions of the photoionization calculations and their parametrization based on a single parameter.

Progress has since been made on the photoionization modeling front. Using a multi-wavelength campaign of NGC 5548, \cite{Mehdipour15} produced a broadband spectrum of an unobscured (Type I) and obscured (Type II) AGN. \cite{Dannen19}, hereafter D19, used these composite SED and an updated version of \textsc{XSTAR}, including the most up-to-date atomic line lists, to compute force multipliers, heating rates and opacities due to spectral lines. They found several key differences with SK90. Firstly, X-ray lines make comparable contributions as UV lines to the force multiplier, both in terms of the number of lines and their oscillator strengths. The upshot is that part of the parameter space near $\log \xi \sim 3$, which was previously thought to be too ionized to allow line driving, in fact, has a force multiplier that can provide a boost of $M(t) \sim 10^{1-2}$ above electron scattering.

In this work, we employ the aforementioned tables of heating and cooling rates and force multipliers as a function of gas ionization parameter, $\xi$, temperature, $T$, and optical depth parameter, $t$, to model AGN disc winds. This represents a significant step forward in modeling line-driven AGN disc winds. Firstly, our hydrodynamic modeling of the line force is based on photoionization modeling that uses the most up-to-date atomic line lists and an observationally motivated irradiating SED. Secondly, our approach does not rely on analytic fits for heating/cooling or force multiplier but rather a robust table interpolation scheme that is more accurate across the surveyed parameter space. Finally, we build on our methods presented in Paper I, where we account for the time-dependent radiation transfer of ionizing X-rays, including scattering and absorption. 

The contents of our paper is as follows. In \S \ref{sec:theory} we describe our simulation set-up, in particular the treatment of the heating and cooling and line opacities, which we compute using the photoionization code \textsc{XSTAR} assuming the flow is irradiated by a composite AGN SED. In \S \ref{sec:results} we describe the results for a suite of simulations exploring the AGN parameter space (black hole mass $M_{\rm{BH}}$, Eddington fraction $\Gamma$ and ionizing photon fraction $f_{\xi}$ and SED hardness). Finally, we conclude in \S \ref{sec:discussion} where we discuss our results in the context of future modeling of AGN line-driven disc winds.

\section{Theory}
\label{sec:theory}
\subsection{Basic Equations}
\label{sec:equations}
The basic equations for single-fluid radiation hydrodynamics are
\begin{subequations}
\begin{equation}
\frac{\partial \rho}{\partial t} + \nabla \cdot \left( \rho \mathbf{v} \right) = 0,
\end{equation}
\begin{equation}
\frac{\partial (\rho \mathbf{v})}{\partial t} + \nabla \cdot \left(\rho \mathbf{vv} + \sf{P} \right) =  \mathbf{G} + \rho \mathbf{g}_{\rm{grav}},
\label{eq:momentum}
\end{equation}
\begin{equation}
\frac{\partial E}{\partial t} + \nabla \cdot \left( (E + P)\mathbf{v} \right) = cG^{0} + \rho \mathbf{v} \cdot \mathbf{g}_{\rm{grav}} + \rho \mathcal{L},
\label{eq:energy}
\end{equation}
\label{eq:hydro}
\end{subequations}
\noindent where $\rho$ is the fluid density, $\mathbf{v}$ the velocity and $\sf{P}$ a diagonal tensor with components $P$ the gas pressure. The total gas energy density is $E = \frac{1}{2} \rho |\mathbf{v}|^2 + \mathcal{E}$ where $\mathcal{E} =  P/(\gamma -1)$ is the internal energy density and $\gamma$ the adiabatic index. The radiation momentum and energy source terms are $\mathbf{G}$ and $G^0$, respectively. We break up the contributions to the radiation field into two parts, an ionizing component and a line-driving component, which we describe more fully in Section \ref{sec:radTransfer}. $\mathcal{L}$ is an optically thin radiative heating term (see \S \ref{sec:photoionization}) and $\mathbf{g}_{\rm{grav}}$ is the gravitational acceleration. Our simulation domain has been shifted upwards in $z$, so $\theta = \pi/2$ corresponds to the disc photosphere, so the gravitational potential must be shifted accordingly (see Paper I). 

The temperature is $T = (\gamma -1)\mathcal{E}\mu m_{\rm{p}}/\rho k_{\rm{b}}$ where $\mu$ is the mean molecular weight, $m_p$ the proton mass and $k_b$ the Boltzmann constant. 

\subsection{Radiation Transfer}
\label{sec:radTransfer}

\begin{figure}
    \centering
    \includegraphics[scale=0.52]{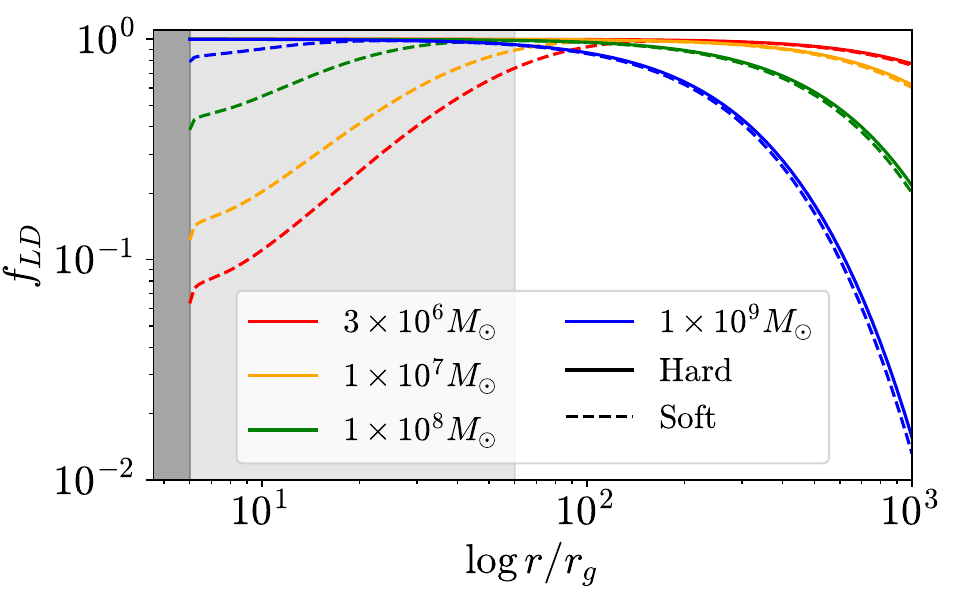}\caption{Radiation field along the disc for central black hole masses $M_{BH}/ M_{\odot} = 3 \times 10^6$ (red), $10^7$ (orange), $10^8$ (green), $10^9$ (blue). The dark grey shading shows radii interior to the ISCO and the light grey shading shows radii interior to the hydro simulation grid. Line driving radiation fraction $f_{LD}$ as a function of radial position for line driving cutoff wavelength $\lambda_1 = 200 $\AA $\:$(soft SED, dashed line) and 1 \AA $\:$ (hard SED, solid line). Neglecting the line driving contribution from radiation in the extreme UV/soft X-ray  1\AA $\leq \lambda \leq 200 $\AA  $\:$ range significantly weakens the line driving strength in the innermost parts of the disc, particularly in the case of low $M \lesssim 10^7 M_{\odot}$ black holes. In the case of a $\lambda_1 = 200$\AA $\:$ cutoff, we see the disc LD fraction is a monotonic function of black hole mass in the inner region, whereas with the $\lambda_1 = 1$\AA $\:$ cutoff $f_{LD} \approx 1$.}
    \label{fig:rad_field}
\end{figure}

The radiation source terms $\mathbf{G}$ and $cG^0$ are assumed to receive contributions from a line driving component (LD) and an ionizing component ($\xi$)
\begin{subequations}
\begin{equation}
\mathbf{G} = \mathbf{G}_{\rm{LD}} + \mathbf{G}_{\xi}, 
\end{equation}
\begin{equation}
G^{0} = G^{0}_{\rm{LD}} + G^{0}_{\xi}. 
\end{equation}    
\end{subequations}
We colloquially think of the UV band as the main contributor to line driving and the X-ray band as the main contributor to gas ionization. This is also in line with the notation of PK04 and Paper I. Of course, this band separation is not strictly correct: X-ray photons do provide momentum transfer to the gas via line scattering, and UV photons above 13.6 eV serve to ionize the gas.

Effectively, our model approximates the system as a compact central source illuminating the gas with ionizing radiation (the $\xi$ component) and emission from the accretion disk (the LD component), imparting momentum on the gas via line opacity. 

We model the ionizing central source radiation ($\xi$ component) via the radiation hydrodynamics module in \textsc{Athena++} by directly solving the gray (frequency averaged) time-dependent radiation transport equation using the implicit radiation solver (see \cite{Jiang2021} for details of the numerical algorithm and Paper I for a more complete description of the numerical setup as applied to disc winds.). The ionizing component is assumed to be incident from the inner radial boundary. The gas opacity is assumed to be due to electron scattering, with a scattering and absorption opacity $\kappa_s = \kappa_a = \kappa_{es}$. The intensity of the selected rays is chosen such that the total ionizing radiation flux is a fraction $f_{\xi}$ of the disc Eddington fraction. We vary this parameter in different models to study the effects of ionization on the outflows.

We model the disc emission (LD component) as time-independent and the wind is assumed to be optically thin to this part of the continuum. The momentum transfer can be broken up into contributions from electron scattering and radiation pressure on spectral lines, 
\begin{equation}
\mathbf{G}_{\rm{LD}} = \frac{\rho \kappa_{\rm es}}{c} \int \mathbf{n} I_D f_{LD} \left[ 1 + M(t) \right] d\Omega,
\end{equation}
where the surface integral is over the disc, assumed to be radiating like a self-irradiated Shakura-Sunyaev disc \citep{SS73} with a constant accretion rate $\dot{M}$. Even though mass is injected into the domain, there is no feedback on the assumed mass flow rate, which remains constant. Hence, the radiation at the disk surface is modeled as
\begin{dmath}
I_D = \frac{3}{\pi}\frac{GM_{BH}}{r_{*}^2}\frac{c}{\kappa_{\rm es}} \Gamma_D \left[ \left(\frac{r_{*}}{r}\right)^3 \left( 1 - \left[\frac{r_{*}}{r}\right]^{1/2} \right) + \frac{x}{3\pi} \left\{ \sin^{-1}\left( \frac{r_{*}}{r} \right) - \frac{r_{*}}{r} \left( 1 - \left[ \frac{r_{*}}{r}\right]^2 \right)^{1/2} \right\} \right],
\label{eq:SSIntensity}
\end{dmath}
where $r_*=6 GM/c^2$, $x$ is the re-radiation factor ($x=1$ in this work), and $\Gamma_D = \dot{M}/\dot{M}_{\rm Edd}$, where
\begin{equation}
\dot{M}_{\rm Edd} = \frac{4 \pi G M_{BH}}{\eta \kappa_{\rm es} c}.
\end{equation}
We assume an efficiency $\eta=1/12$ in this work. In this model, the spectrum is blackbody, with a temperature
\begin{equation}
    T_D = \left[ \frac{\pi I_D}{\sigma} \right]^{1/4}.
    \label{eq:SSTemp}
\end{equation}
The fraction of the intensity in the band that contributes to line driving is then found by integrating the Planck function 
\begin{equation}
    f_{\rm{LD}} = \frac{\pi}{\sigma T_D^4} \int_{\lambda_1}^{\lambda_2} \frac{2hc^2}{\lambda^5} \frac{1}{ \exp \left\{hc / \lambda k_b T_D \right\}- 1},
\end{equation}
with $\sigma$ is the Stefan-Boltzmann constant. 
For most of this work, we take the contributions to the line driving as a combination of UV and X-ray photons in the range $\lambda_1 = 1$ \AA \ and $\lambda_2 = 3 200$ \AA. This choice was motivated by the results of the photoionization study, D19, which found that X-ray bands have significant numbers of strong lines and thus contribute to the line force. We further motivate this choice in Sec \ref{sec:photoionization}. This differs from previous works (PK04 and Paper I), which assumed only UV photons contributed to line driving and hence used cutoffs $\lambda_1 = 200$ \AA \ and $\lambda_2 = 3 200$ \AA. In Fig \ref{fig:rad_field} we plot $f_{LD}$ as a function of radial distance for black hole masses $M_{BH}/ M_{\odot} =$ $3 \times 10^6$ (red), $10^7$ (orange), $10^8$ (green), $10^9$ (blue). The dark grey shading shows radii interior to the ISCO and the light grey shading radii interior to the inner radius of the simulation grid. The solid lines are for wavelength cutoff $\lambda_1 = 1$ \AA \ and the dashed lines for $\lambda_1 = 200$ \AA. We will refer to these as the \emph{hard} and \emph{soft} SED, respectively. The soft SED  has a significantly weaker line force in the innermost parts of the disc, particularly in the case of low $M \lesssim 10^7 M_{\odot}$ black hole mass. For the soft SED, the $f_{\rm{LD}}$ is a monotonic function of black hole mass, whereas for the hard SED, it is approximately $f_{\rm{LD}} \approx 1$ across the whole mass scale. We study the effects of changing the UV cutoff in Section \ref{sec:SEDEffects}.

The strength of the spectral lines, relative to electron scattering, is quantified via the force multiplier $M(t,\xi)$ (\cite{Owocki1988}). The force multiplier is a function of the gas ionization $\xi$ and the optical depth parameter 
\begin{equation}
t = \frac{\kappa_{\rm es} \rho v_{\rm{th}}}{\left| dv_{l} / dl\right|},
\end{equation}
with the thermal velocity of the gas $v_{\rm{th}} = 4.2 \times 10^5$ cm/s, which is a measure of the total number of available lines. We compute the force multiplier contribution at each hydrodynamic grid zone by using bilinear extrapolation along precomputed tables for the force multiplier as a function of these two parameters. We describe the generation of these tables more fully in \S \ref{sec:photoionization}. 

The key difference between this work and Paper I is using pre-computed tables for the heating and line force generated using the photoionization code \textsc{XSTAR}. We use this setup to study the effects of varying disc Eddington fraction, black hole mass and ionizing radiation fraction on the resulting disc winds.

\subsection{Photoionization Modeling}
\label{sec:photoionization}

We employ the photoionization code \textsc{XSTAR} to compute a grid of models to implement microphysics into our disc wind hydro simulations. For a wide range of temperature and photoionization parameter, we compute the force multiplier (see D19 for a full description of our methods) and heating/cooling (see \cite{Dyda17}). This grid of models is also used in our interpolation function for our heating and cooling scheme described in many of our previous studies of thermally driven winds \citep[e.g., see][]{Dannen20,Ganguly21}.  

\subsubsection{Force Multiplier}

\begin{figure*}
    \centering
    \includegraphics[scale=0.56]{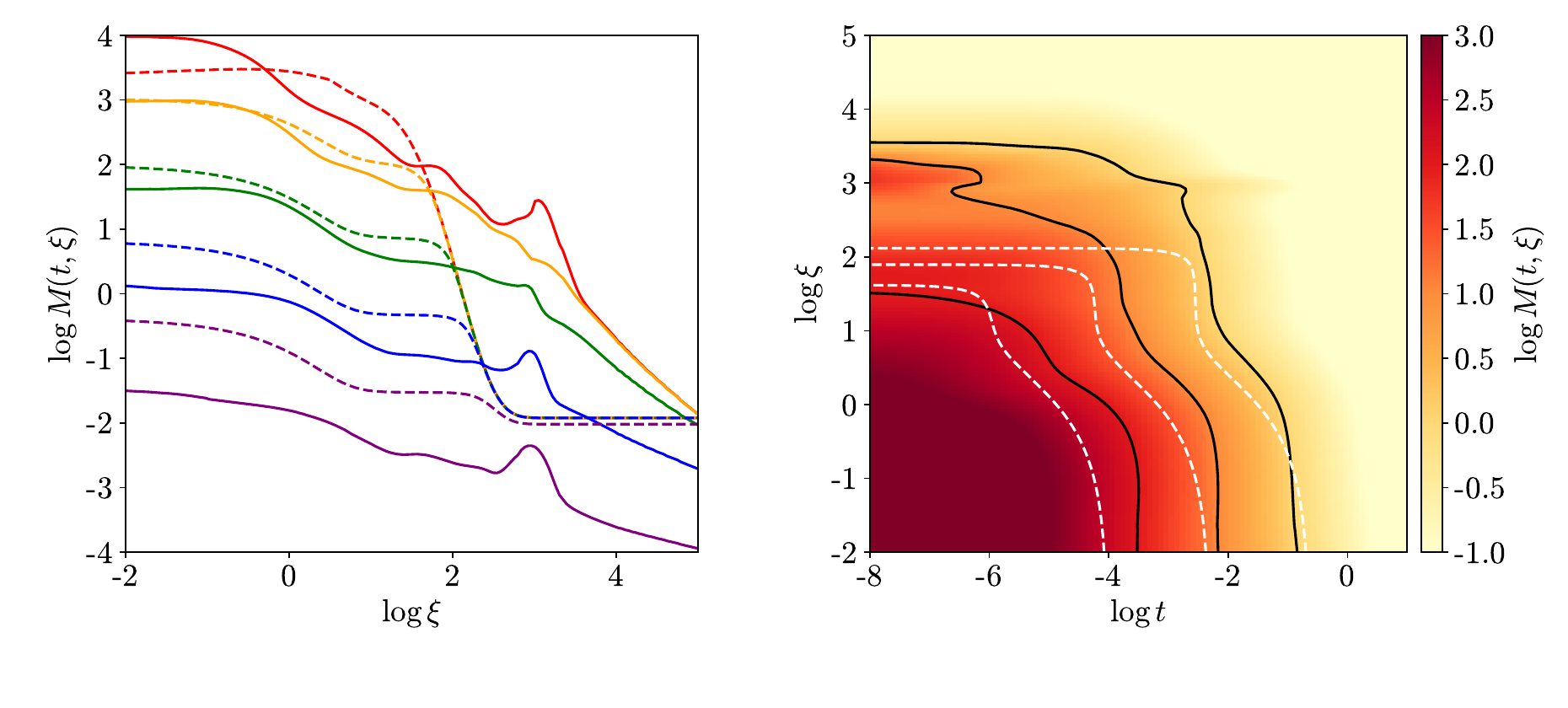}
    \caption{Comparison of force multiplier computed using our \textsc{XSTAR} photoionization calculation (solid lines) and the analytic expression (dashed lines). \textit{Left panel -} Force multiplier as a function of ionization parameter $\xi$. Each lines color is for fixed optical depth parameter $\log t = $ $ 10^{-8}$ (red), $ 10^{-6}$ (orange), $ 10^{-4}$ (green), $ 10^{-2}$ (blue) and $ 10^{0}$ (purple). \textit{Right panel -} Contour plot of force multiplier in the ionization parameter and optical depth parameter space. The three solid black lines show the $\log M(t,\xi) = 0,1,2$ contours. The dashed white lines show the same contours for the analytic expression as in Paper I. This plot clearly demonstrates the upper, left region of parameter space where the force multiplier is strong, $0 \leq \log M(t,\xi) \leq 1$, but for which the old analytic expression predicted it was weak, $\log M(t,\xi) \lesssim 0$.}
    \label{fig:AGN_Mt}
\end{figure*}
The radiation force due to spectral lines is characterized by the force multiplier $M(t)$, which is the strength of the radiation force relative to electron scattering. The force multiplier prescription used in previous disc wind studies (PSK00, PK04, Paper I) used a modified CAK formulation (\cite{Owocki1988}) that relied on analytic fits to the photoionization studies of SK90. We used this analytic model in Paper I (see equations (18) - (20)).

D19 used an updated version of \textsc{XSTAR}, including a revised line list featuring over 2 million lines. This primarily includes the \cite{Kurucz1995} line list, supplemented by the CHIANTI X-ray line list \citep{CHIANTIMethod,CHIANTI2020}. D19 found that contrary to SK90, which used the \cite{Abbott1982} line list, the number and quality of lines (as measured by oscillator strength) do not decrease precipitously at higher energies and ionization degrees, as shown in their Fig. 2.

We illustrate the effect of these higher energy lines in Fig \ref{fig:AGN_Mt} where we plot the force multiplier computed using D19 photoionization calculations (solid lines) and the analytic expression used in Paper I (dashed lines). The left panel shows the force multiplier as a function of ionization parameter $\xi$. The colored lines are for fixed optical depth parameter $\log t = $ $-8$ (red), $-6$ (orange), $-4$ (green), $-2$ (blue) and $0$ (purple).

For wind to launch, the outward radiation force must overcome the inwards force of gravity, hence $\Gamma_D M(t,\xi) \gtrsim 1$. The \textsc{XSTAR} calculations find such a condition is met for $2.5 \leq \log \xi \leq 3.5$ and $-8 \leq \log t \leq -4$ whereas the analytic expression does not. The right panel shows the contour plot of the force multiplier in the ionization parameter and optical depth parameter space. The three solid black lines show the $\log M(t,\xi) = 0,1,2$ contours. The dashed white lines show the same contours for the analytic expression in Paper I. This plot clearly demonstrates the upper, left region of parameter space where the force multiplier is strong, $0 \leq \log M(t,\xi) \leq 1$, but for which the old analytic expression predicted it was weak, $\log M \lesssim 0$.  

As a result of these findings, we have modified the line-driving prescription used in previous works 1) We compute the force multiplier by interpolating the D19 force multiplier tables rather than using an analytic expression. 2) We include extreme UV and soft X-rays to the band contributing to line driving, decreasing the wavelength cutoff from $\lambda_1 = 200$ \AA \ to 1 \AA. 

\subsubsection{Heating Prescription}
\label{sec:heating}

\begin{figure*}
    \centering
    \includegraphics[scale=0.56]{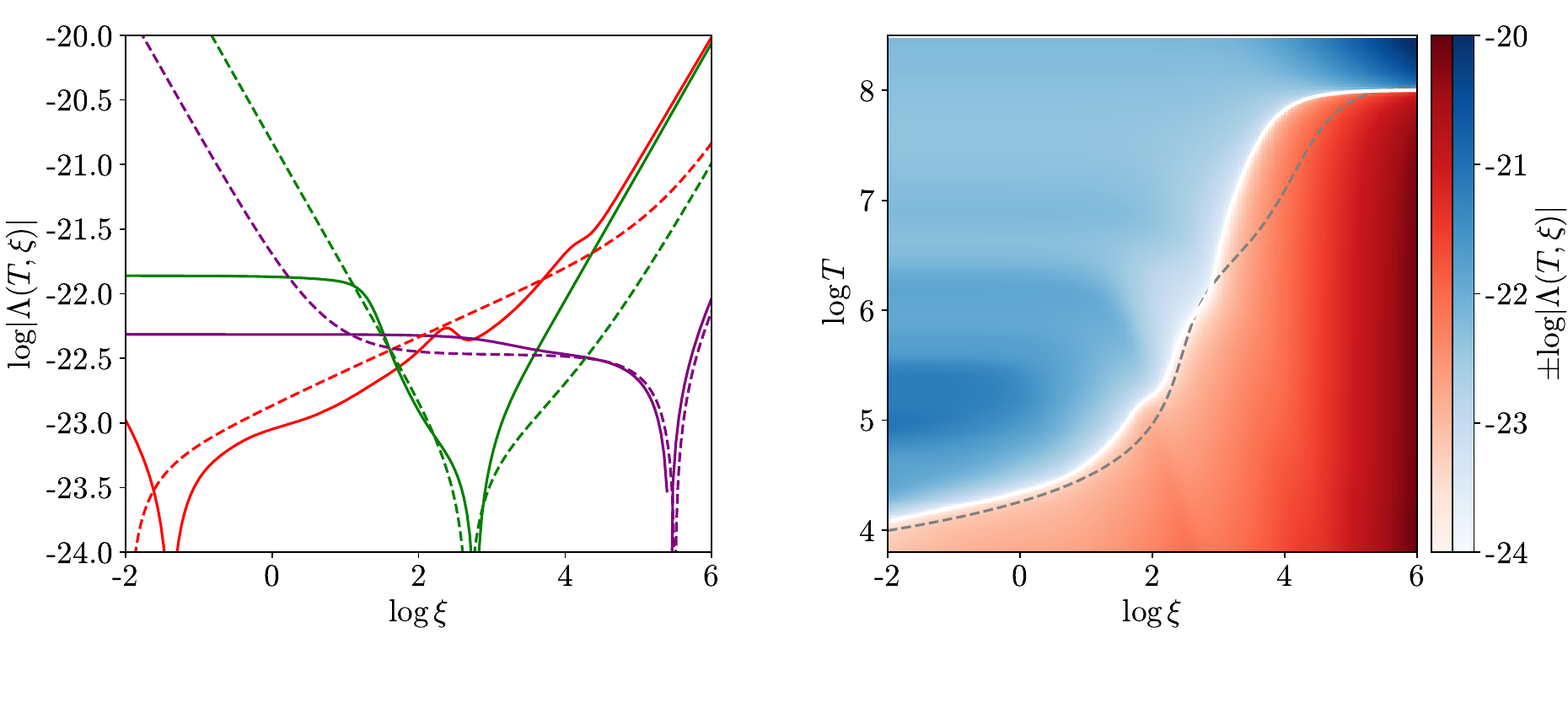}
    \caption{Heating and cooling  computed from \textsc{XSTAR} photoionization calculation. \textit{Left panel -} Heating and cooling as a function of ionization parameter $\xi$ for AGN SED (solid lines) and analytic (dashed lines). The colored lines are for fixed temperature $\log T = $ $4$ (red), $6$ (green) and $8$ (purple). \textit{Right panel -} Heating (red shade) and cooling (blue shade) contour map as a function of temperature and ionization parameter for AGN SED. The equilibrium S-curve (white line) is well resolved in our parameter space. We overplot the analytic S-curve (grey dashed line), which agrees at high temperatures where Compton processes dominate but differ at intermediate temperatures and ionizations due to free-free and line processes.}
    \label{fig:AGN_HC}
\end{figure*}

We assume locally optically thin radiative heating and cooling due to Compton, X-ray photoionization, Bremsstrahlung, and atomic line processes, assuming a 10keV Bremsstrahlung incident SED. Previous disc wind models used a modified form of the heating in SK90 and \cite{Blondin1994}, which was a semi-analytic fit to the heating found using their photoionization model (see their appendix for analytic expression). 

We computed the heating/cooling rate using \textsc{XSTAR} assuming a gas irradiated by the composite type I AGN spectra in \cite{Mehdipour15}. In Fig \ref{fig:AGN_HC} we plot the net heating/cooling of this model for gas irradiated by an AGN SED (solid lines) and for the analytic model (dashed lines). The left panel shows the heating and cooling as a function of ionization parameter $\xi$. The colored lines are for fixed temperature $\log T = $ $4$ (red), $6$ (green) and $8$ (purple). The right panel shows the heating (red shade) and cooling (blue shade) contour map as a function of temperature and ionization parameter. We note the equilibrium S-curve (white line) is well resolved in our parameter space. We also overplot the analytic heating S-curve (grey dashed line). The equilibrium curves nearly agree at low and high temperatures but differ in the intermediate range where line effects are dominant.

Studies of spherically symmetric, thermal winds have shown that the shape of the equilibrium curve determines the path through parameter space that the outflow follows. In particular, prior to the critical point, the solution effectively follows the S-curve. \cite{Dyda17} showed explicitly that gas irradiation by the AGN SED or heated via an analytically defined function generates qualitatively different outflows. In a line-driven wind, thermal effects dominate at the base of the wind where the line force is relatively weak. For this reason, and also for self-consistency with our prescription of the line force, we compute the gas heating by interpolating along the heating tables generated from the D19 photoionization model. 

\subsection{Simulation Parameters}
We consider AGN systems with a range of black hole and disc parameters. For ease of comparison between models, we will express results as much as possible in terms of dimensionless parameters. We consider black holes in the mass range $3 \times 10^6 \leq M_{\rm{BH}}/M_{\odot} \leq 10^{9}$. The gravitational radius is then $r_g = GM_{\rm{BH}}/c^2$. The inner edge of the disc is assumed to extend to the innermost stable circular orbit, $r_* = 6 r_g$. We express time in units of the orbital period at the inner radial boundary, $t_0 = 2\pi ((10 r_*)^3/GM)^{1/2}$ with $10 r_*$ the inner radial boundary.

We impose inflow (outflow) boundary conditions at the inner (outer) radial boundaries and axis boundary conditions along the $\theta$ = 0 axis. We assume a reflection symmetry about the $\theta = \pi/2$ midplane.
We use a vacuum boundary condition for the radiation along the disc midplane and outer radial boundaries and keep the ionizing radiation flux fixed at the inner radial boundary. After every full time step we reset $\rho_d = 10^{-8} \rm{g \ cm^{-3}}$, $v_r = 0$ and $v_{\phi} = v_K = \sqrt{GM/r}$. We also impose that the vertical velocity component $v_{\theta}$ is unchanged due
to resetting density.

We choose a domain size $n_r$ × $n_{\theta}$ = 96 × 140. The radial domain extends over the range $10 \ r_{*} < r < 500 \ r_{*}$ with geometric spacing $dr_{i+1}/dr_{i} = 1.05$. The polar angle range is $0 < \theta < \pi/2$ and has geometric spacing $d \theta_{j+1}/d \theta{j} = 0.938$, which ensures that we have sufficient resolution near the disc midplane to resolve the acceleration of the flow.

Initially, the cells along the disc are set to have $\rho = \rho_d$, $v_r = v_{\theta} = 0$, $v_{\phi} = v_K$. In the rest of the domain $\rho = 10^{-20} \ \rm{g \ cm^{-3}}$ and $v_r = v_{\theta} = v_{\phi} = 0$. Everywhere, the temperature is constant along vertical cylinders corresponding to the Shakura-Sunyaev disc temperature at the base given by \cite{SS73}.

The disc Eddington fraction $0.2 \leq \Gamma_D \leq 0.5$ and re-radiation factor $x = 1$ (see eq \ref{eq:SSIntensity}). The fraction of central source luminosity in X-rays is $0.01 \leq f_{\xi} \leq 0.1$ of the corresponding disk luminosity. We assume the central source does not emit photons that contribute to the line force. The source of line-driving photons is assumed to extend all the way from the ISCO (effectively outside the simulation domain) with $r_{*} \leq R_d \leq 1500 r_{*}$. A sphere of radius $r_{*}$, effectively the black hole and nearby corona, is assumed to be optically thick to shield the wind from the backside of the disc.  

We impose a density floor $\rho_{\rm{floor}} = 10^{-22} \rm{g \ cm^{-3}}$ which adds matter to stay above this floor, while conserving momentum, if the density ever drops below it. In addition, we have the temperature floor as a function of cylindrical radius.

\section{Results}
\label{sec:results}

\begin{table*}
\centering
\begin{tabular}{ l c c c c c c c c c l}
\hline \hline
\multirow{2}{*}{Model} & \multirow{2}{*}{$M_{BH}$ \ [$10^8 \ M_{\odot}$]} & \multicolumn{2}{c}{Radiation } & \multicolumn{5}{c}{Wind Properties} & Description \\
& & $\Gamma_d$  & $f_{\xi}$  & $\log \dot{m} \ [M_{\odot}/\rm{yr}]$ & $\log \rho_{\rm{out}} \ [{\rm g\,cm}^{-3}] $ & $v_{\rm{out}} $ [km/s] & $\omega$ & $\Delta \omega$ \\ \hline \hline
GM10G05x005   & 10 & 0.5 & 0.05 & 1.99  & -16.5 & 18 \ 700 & 65 & 10 & Wind \\
GM10G04x005   & 10 & 0.4 & 0.05 & 0.79  & -16.8 & 9 \ 500 & 69 & 9 & Wind \\
GM10G03x005   & 10 & 0.3 & 0.05 &  -- & -- & -- & -- & -- & No Wind \\ \hline
GM1G05x005    & 1  & 0.5 & 0.05 & 1.33  & -15.4 & 24 \ 500 & 65 & 17 & Wind \\
GM1G04x005    & 1  & 0.4 & 0.05 & 1.09  & -15.3 & 15 \ 200 & 66 & 12 & Wind \\
\rowcolor{lightgray}
GM1G03x007    & 1  & 0.3 & 0.07 & --      & --     & --       & --  & --  & No Wind \\
GM1G03x005    & 1  & 0.3 & 0.05 & -0.82  & -16.8 & 7 \ 000 & 66 & 12 & Wind \\  \rowcolor{lightgray}
GM1G03x003    & 1  & 0.3 & 0.03 & -0.77  & -17.5 & 6 \ 000 & 67 & 10 & Wind \\ \rowcolor{lightgray}
GM1G03x001    & 1  & 0.3 & 0.01 & 1.01  & -15.6 & 17 \ 500 & 68 & 19 & Wind \\
GM1G02x005    & 1  & 0.3 & 0.05 & --      & --     & -- & -- & -- & No Wind \\ \hline
GM01G05x005   & 0.1 & 0.5 & 0.05 & 0.47  & -14.2 & 23 \ 000 & 64 & 23 & Wind \\
GM01G04x005   & 0.1 & 0.4 & 0.05 & 0.28  & -14.3 & 21 \ 000 & 62 & 11 & Wind \\
GM01G03x005   & 0.1 & 0.3 & 0.05 & -1.38  & -16.4 & 4 \ 000 & 70 & 10 & Wind \\ \hline
GM0033G05x005 & 0.033 & 0.5 & 0.05 & -- & -- & -- & -- & -- &  Unresolved Wind \\
GM0033G04x005 & 0.033 & 0.4 & 0.05 & -- & -- & -- & -- & -- &  Unresolved Wind \\
\rowcolor{lightgray}
GM0033G03x010 & 0.033 & 0.3 & 0.10 & -- & -- & -- & -- & -- & No Wind \\ \rowcolor{lightgray}
GM0033G03x007 & 0.033 & 0.3 & 0.07 & -2.42  & -16.0 & 4 \ 700  & 67 & 10 & Wind \\
GM0033G03x005 & 0.033 & 0.3 & 0.05 & -2.11  & -15.2 & 7 \ 000 & 66 & 10 & Wind \\ \rowcolor{lightgray}
\rowcolor{lightgray}
GM0033G03x003 & 0.033 & 0.3 & 0.03 & -2.33 & -15.6 & 9 \ 700 & 63 & 12 &  Wind \\ \rowcolor{lightgray}
GM0033G03x001 & 0.033 & 0.3 & 0.01 & -0.38  & -15.0 & 19 \ 100  & 57 & 16 &  Wind \\
GM0033G02x005  & 0.033 & 0.2 & 0.05 & -- & -- & -- & -- & -- & No Wind \\
\hline\hline
GM10G05x005$\lambda$ & 10 & 0.5 & 0.05 & 1.63  & -17.0 & 24 \ 500 & 67 & 14 & Wind \\
GM10G04x005$\lambda$ & 10 & 0.4 & 0.05 & -1.36 & -17.6 & 12 \ 500 & 64 & 10 & Wind \\
GM10G03x005$\lambda$ & 10 & 0.3 & 0.05 & -- & -- & -- & -- & -- & No Wind \\ \hline
GM1G05x005$\lambda$  & 1  & 0.5 & 0.05 & 1.15  & -15.8 & 31 \ 700 & 60 & 20 & Wind \\
GM1G04x005$\lambda$  & 1  & 0.4 & 0.05 & 1.17  & -15.0 & 19 \ 400 & 66 & 31 & Wind \\
GM1G03x005$\lambda$  & 1  & 0.3 & 0.05 & -1.17  & -16.5 & 4 \ 800 & 71 &  6 & Wind \\ \hline
GM01G05x005$\lambda$ & 0.1 & 0.5 & 0.05 & 0.44  & -14.0 & 19 \ 000 & 65 & 32 & Wind \\
GM01G04x005$\lambda$ & 0.1 & 0.4 & 0.05 & 0.29  & -14.0 & 18 \ 600 & 66 & 37 & Wind \\
GM01G03x005$\lambda$ & 0.1 & 0.3 & 0.05 & -- & -- & -- & -- & -- & No Wind \\ \hline
GM0033G03x005$\lambda$ & 0.033 & 0.3 & 0.05 & -5.03  & -18.0 & 1 \ 200 & 67 & 15 & Wind \\ \hline \hline
    \end{tabular}
\caption{Summary of models listing the black hole mass $M_{BH}$, UV Eddington fraction $\Gamma_{d}$, ionizing Eddington fraction $f_{\xi}$. Models ending with a $\lambda$ use a softer cutoff for the UV distribution $\lambda_1 = 200 \AA$. We list the time average wind properties for models with an outflow, including the total wind mass flux $\dot{m}$, the typical wind density $\rho_{\rm{out}}$ and velocity$v_{\rm{out}}$ at the outer boundary, as well as the wind angle with the zenith $\omega$ and opening angle $\Delta \omega$. The shading indicates the models where $f_{\xi}$ is varied to study the effects of ionization.}
\label{tab:summary}
\end{table*}

\begin{figure*}
    \centering
    \includegraphics[scale=1.1]{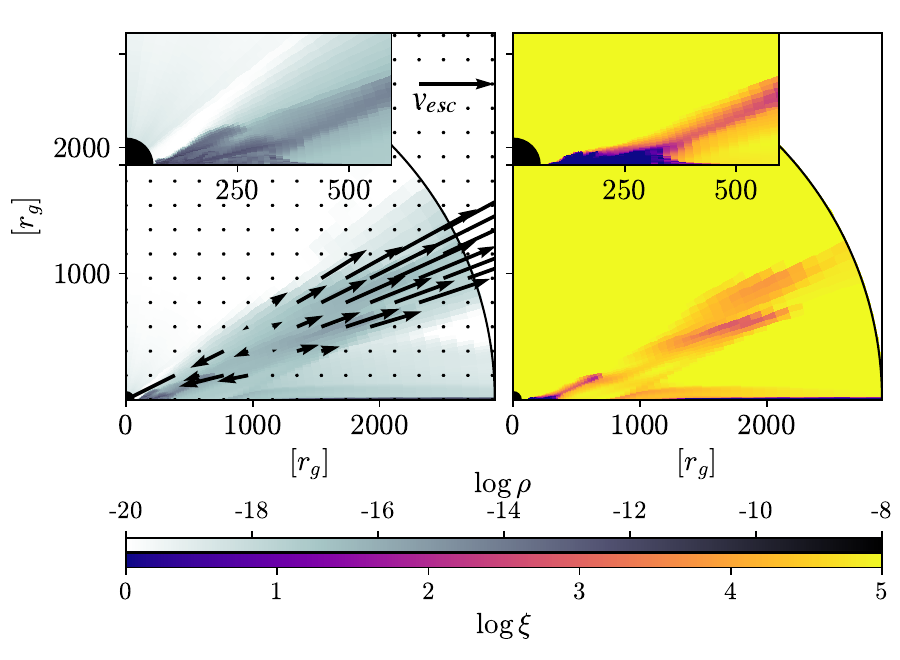}
    \caption{Typical outflow structure for AGN disc wind, time averaged over 5 inner disc orbits. We include a zoom-in panel within the first $500 \ r_g$ near the black hole. \textit{Left panel -} Wind density and velocity, showing that the outflow is somewhat episodic, with some gas close (inside of $r \lesssim 1000 r_g$) to the black hole falling inwards, yet gas further out near $r \sim 2000 r_g$ being launched outwards at faster than the escape velocity. \textit{Right panel} - Ionization parameter of the flow structure. Line driving can overcome gravity when $\log \xi \lesssim 3$, such as at radii $r \sim 1500 r_g$ where the gas accelerates to escape velocity, and at $r \lesssim 500 r_g$ where the density is high due to a failed wind.}
    \label{fig:2dDensityXipanel}
\end{figure*}

Below, we describe the results of a series of AGN disc wind simulations. We primarily explored the effects of varying black hole mass, disc accretion rate/Eddington fraction and the ionizing radiation fraction. We also studied the effects of varying the part of the spectra that accounts for the line driving. A complete list of simulations and the basic parameters are listed in Table \ref{tab:summary}.  

We observed two qualitatively different simulation behaviours. Firstly, for some parameter combinations, no significant outflows were launched. After an initial transient period where some low-density material is lifted up above the disc, most gas settles into a hydrostatic atmosphere above the midplane. For these parameter choices, the radiation force is too weak to overcome gravity. When the disc intensity is too low, $\Gamma_d \lesssim 0.2$, no wind can launch. Alternatively, if $\Gamma_d \gtrsim 0.3$ the wind is suppressed if the fraction of ionizing radiation is too high. In such models, the effective force multiplier, that is to say, the ratio of the radiation force due to lines over the radiation force due to electron scattering, is much less than order unity. The ionizing fraction required to fully suppress the wind depends on the black hole mass and Eddington fraction but is approximately $f_{\xi} \sim 10 \%$. 

For simulations with outflows, the winds were all qualitatively similar. Outflows were approximately radial and launched at an angle $\omega \sim 60^{\circ}$, with an opening angle $\Delta \omega \sim 10^{\circ}$. Wind densities at the outer boundary at $r = 3000 \ r_g$ were in the range $\rho \sim 10^{-(14-16)} \; \rm{g/cm^3}$ and with velocities $v \sim 10-20 \times 10^{3}$ km/s.

We show a typical wind solution in Fig \ref{fig:2dDensityXipanel} with the wind density and velocity (left panel) and ionization parameter (right panel) time averaged over 5 inner disc orbits. We include a zoom-in panel within the first $500 \ r_g$ near the black hole. We note that the outflow is somewhat episodic, with some gas close to the central region, $r \lesssim 1000 \ r_g$, falling inwards, yet gas further out near $r \gtrsim 1500 \ r_g$ accelerating outwards and being launched at faster than the escape velocity. This failed wind increases the density close to the black hole and decreases the ionization parameter, thereby allowing some of it to launch later due to the resulting enhanced force multiplier. It also shields matter farther out, similarly decreasing the ionization parameter and enhancing the line force (see the right panel Fig \ref{fig:2dDensityXipanel}).

\subsection{Black Hole Mass}

\begin{figure}
    \centering
  \includegraphics[scale=0.42]{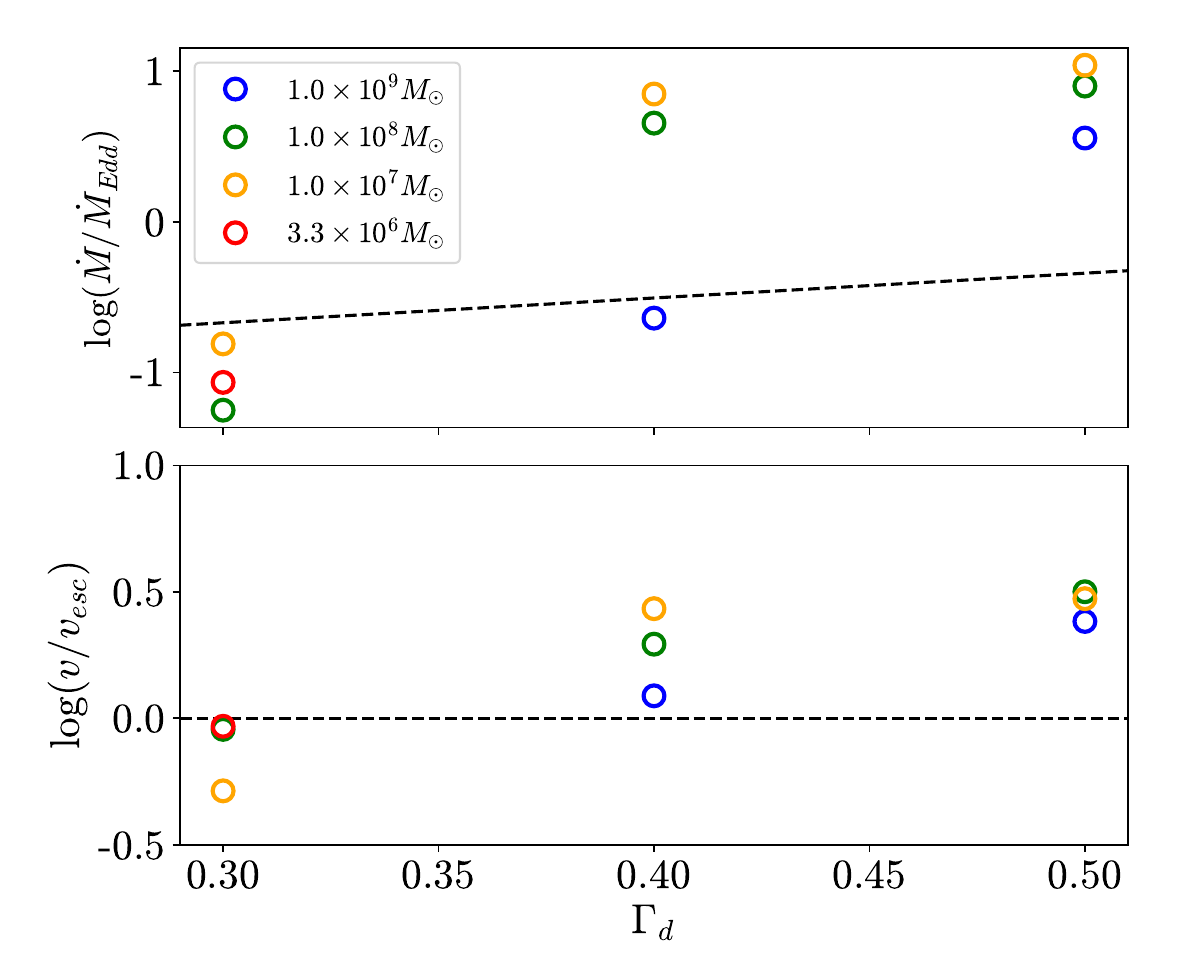}     
    \caption{Wind mass flux in units of Eddington rate (top panel) as a function of disc Eddington fraction for black hole masses $M/M_{\odot} = $ $3.3 \times 10^6$ (red), $1. \times 10^7$ (orange), $1. \times 10^8$ (green) and $1. \times 10^9$ (blue) and ionization $f_{\xi} = 0.05$. The dashed line shows the assumed disc accretion rate. For $\Gamma_{d} > 0.4$ the wind mass flux exceeds the assumed disc accretion rate, indicating our model without wind feedback on the disc is breaking down. For $\Gamma_{d} \approx 0.3$ the mass flux scales is approximately constant in units of the Eddington rate.}
    \label{fig:mdot_scaled}
\end{figure}

\begin{figure}
    \centering
    \includegraphics[scale=0.48]{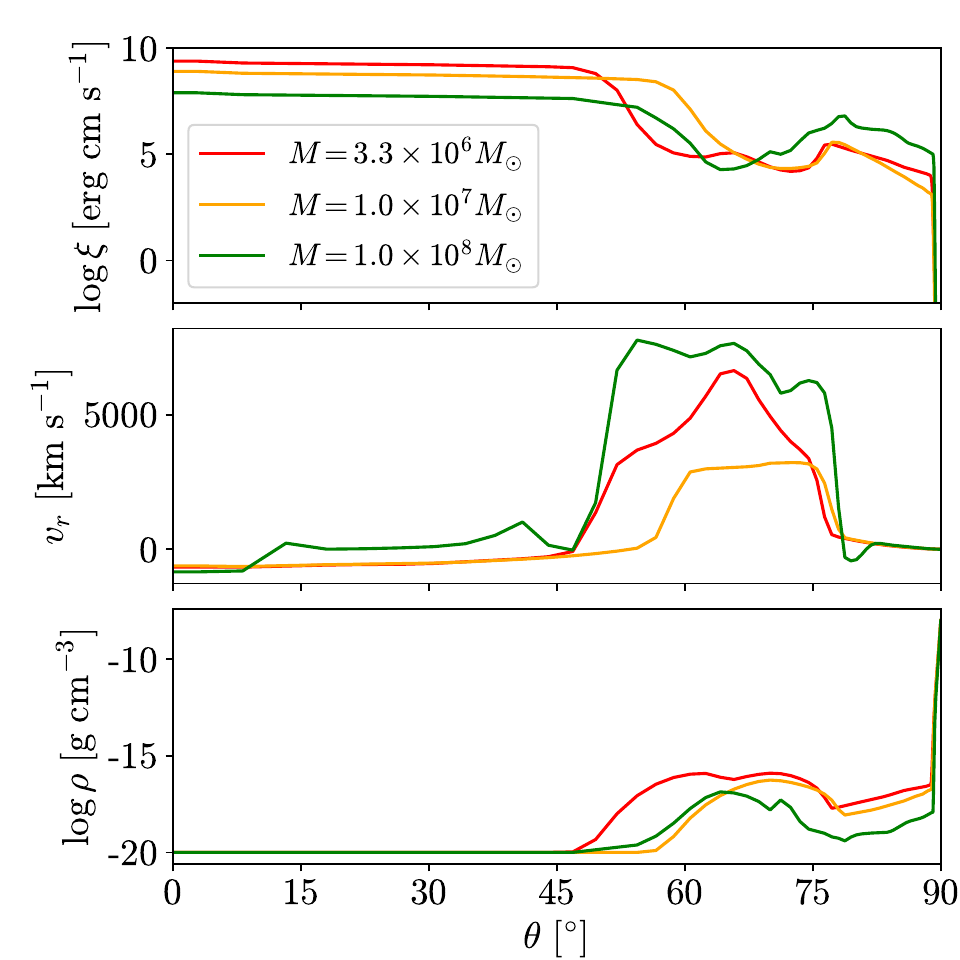}
    \caption{Ionization parameter (top panel), radial velocity (center panel) and density (bottom panel) as a function of altitude at the outer boundary for models with $M = 3.3 \times 10^6 M_{\odot}$ (red), $M = 1.0 \times 10^7 M_{\odot}$ (orange) and $M = 1.0 \times 10^8 M_{\odot}$ (green) and $\Gamma_{d} = 0.3$ and $f_{\xi} = $ 0.05. The ionization is roughly constant in the outflow with $\log \xi \approx 5$. The velocity and density plots show the outflows are concentrated in a narrow region of $\Delta \omega \approx 10^{\circ}$.}
    \label{fig:mass_boundary}
\end{figure}

We perform a study of the effect of black hole mass on the strength of AGN outflows. In Fig. \ref{fig:mdot_scaled}, we plot the time-averaged mass flux through the outer radial boundary (top panel) and outflow velocity (bottom panel) for black hole masses $M/M_{\odot} = $ $3.3 \times 10^6$ (red), $1. \times 10^7$ (orange), $1. \times 10^8$ (green) and $1. \times 10^9$ (blue) and $f_{\xi} = 0.05$. The dashed line shows the assumed disc accretion rate. 

For models with $\Gamma_{d} > 0.4$ the outflow rate is higher than the assumed inflow rate, indicating that the model assumptions of a disc with stationary accretion rate is inconsistent. For the lowest mass cases with $M/M_{\odot} = 3.3 \times 10^{6}$, the acceleration is actually so great that we do not resolve the acceleration zone near the disc. We, therefore, focus our efforts on lower accretion rate cases and discuss possible prescriptions for accounting for feedback of the wind on the disc in Sec \ref{sec:discussion}.

For models with $\Gamma_{d} = 0.3$, the mass outflow is approximately constant, up to a factor of a few, in units of the Eddington rate for black hole masses $3.3 \times 10^{6} \leq M/M_{\odot} \leq 10^{8}$. For $M = 10^{9} M_{\odot}$ no outflow was observed. The mass outflow rates can be understood from the line driving radiation fraction plot in Fig. \ref{fig:rad_field}. All parameters relevant to wind launching, i.e., the gravitational force proportional to $-GM$ and radiation force proportional to $+\Gamma GM$ scale with the black hole mass, therefore any difference in the radiation field, is due solely to changes in $f_{LD}$. For the lower mass models, in the innermost part of the disc, which is responsible for most of the driving, $0.9 \leq f_{LD} \leq 1.0$. Thus, these models exhibit very similar mass outflow rates. For the largest mass case, $M = 10^{9} M_{\odot}$, the UV fraction $f_{LD} \gtrsim 0.5$, leading to a loss in radiative force and ultimately being unable to launch an outflow. This is consistent with the basic theory of line-driven winds, which predicts a mass loss rate that scales with the Eddington fraction. Likewise, the terminal velocity scales with the escape velocity.

For lower Eddington fractions, $\Gamma_{d} \leq 0.2$ (not plotted) outflows are unable to launch across the entire mass scale. This supports our finding that the $M = 10^{9} M_{\odot}$, $\Gamma_{d} = 0.3$ case was unable to launch a wind. Near the minimum disc luminosity, the mass loss rate varies strongly with the Eddington fraction, and small changes in the line force due to a loss of line-driving photons can cause the wind to fail. 

In the lower panel of Fig. \ref{fig:mdot_scaled}, we plot the time-averaged outflow velocity, scaled to the Keplerian velocity at the inner simulation boundary. We do not see a clear trend with black hole mass. However, we see a strong trend with the disc Eddington fraction, with the outflow velocity trending to zero asymptotically as we approach the critical Eddington fraction $\Gamma_{d} \lesssim 0.3$. 

To further understand these trends, we plot the ionization parameter, velocity and density of the wind as a function of altitude along the radial outer boundary in Fig \ref{fig:mass_boundary} for models with $\Gamma_{d} = 0.3$. The colors again correspond to black hole masses of $M/M_{\odot} = $ $3.3 \times 10^6$ (red), $1. \times 10^7$ (orange), $1. \times 10^8$ (green). The ionization structure in the wind very nearly matches with $\log \xi \approx 5 $, slightly above the ionization needed for ideal line driving conditions. Closer to the disc, the $GM = 10^{8} M_{\odot}$ model is more strongly ionized due to a lower gas density. The velocity does not show a clear trend with black hole mass. However, we note that the models with higher mass flux rates can attribute this to a higher gas density and not due to a higher outflow velocity. Likewise, the velocity and density plots show that the angular size of the outflow is rather narrow with $\Delta \omega \sim 10^{\circ}$ as either the density of velocity drops sharply from its central value.

\subsection{Ionization Effects}

\begin{figure}
    \centering
  \includegraphics[scale=0.42]{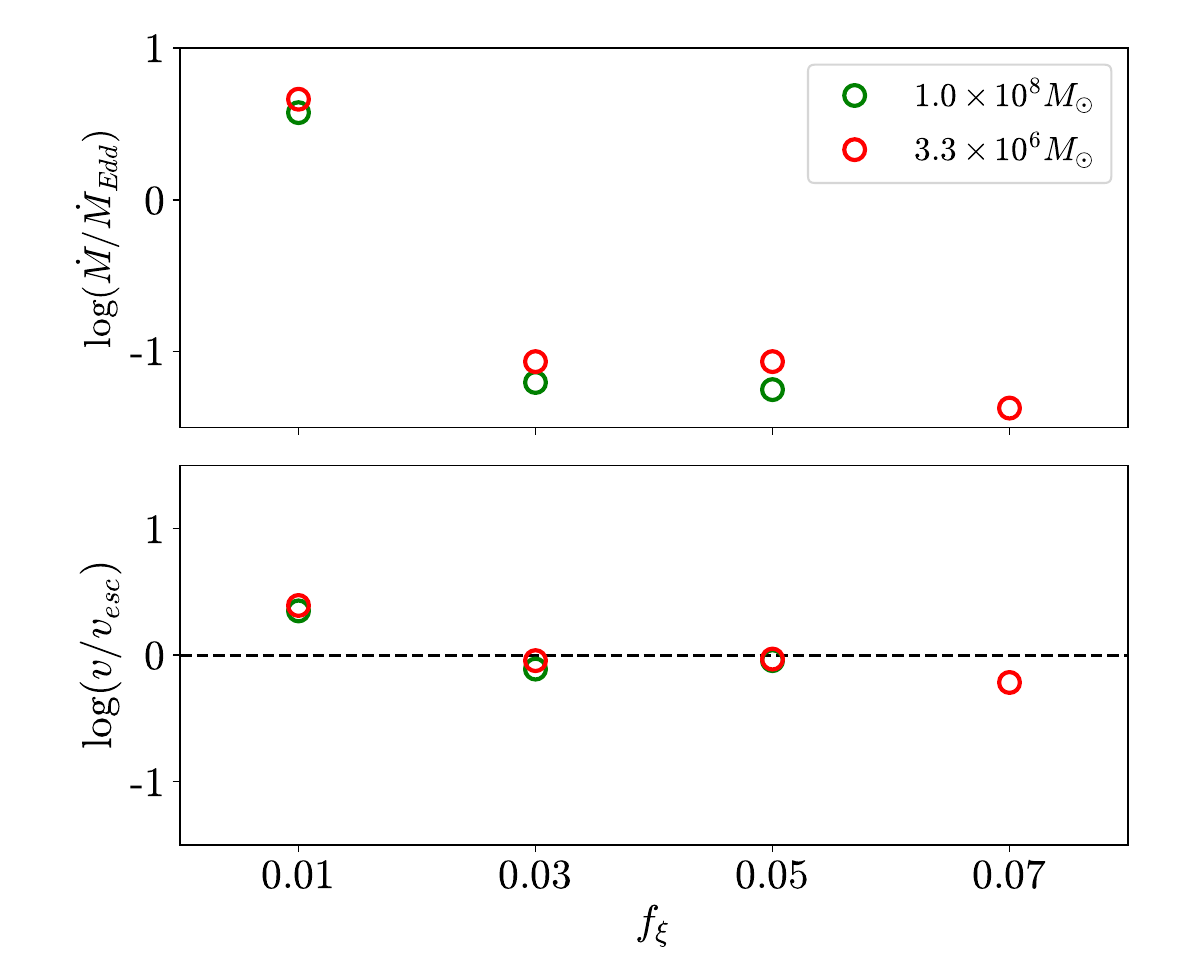}
    \caption{Wind mass flux in units of Eddington rate and outflow velocity as a function of ionizing Eddington fraction for black hole masses $M/M_{\odot} = $ $3.3 \times 10^6$ (red) and $1. \times 10^8$ (green). At low levels of ionization, $f_{\xi} \lesssim 0.01$ outflows are enhanced both in terms of mass flux and outflow velocity. These are both roughly constant at moderate ionizations until the outflow is suppressed beyond an ionization threshold $f_{\xi} \gtrsim 0.1$.}
    \label{fig:mdot_ionization}
\end{figure}

\begin{figure}
    \centering
    \includegraphics[scale=0.48]{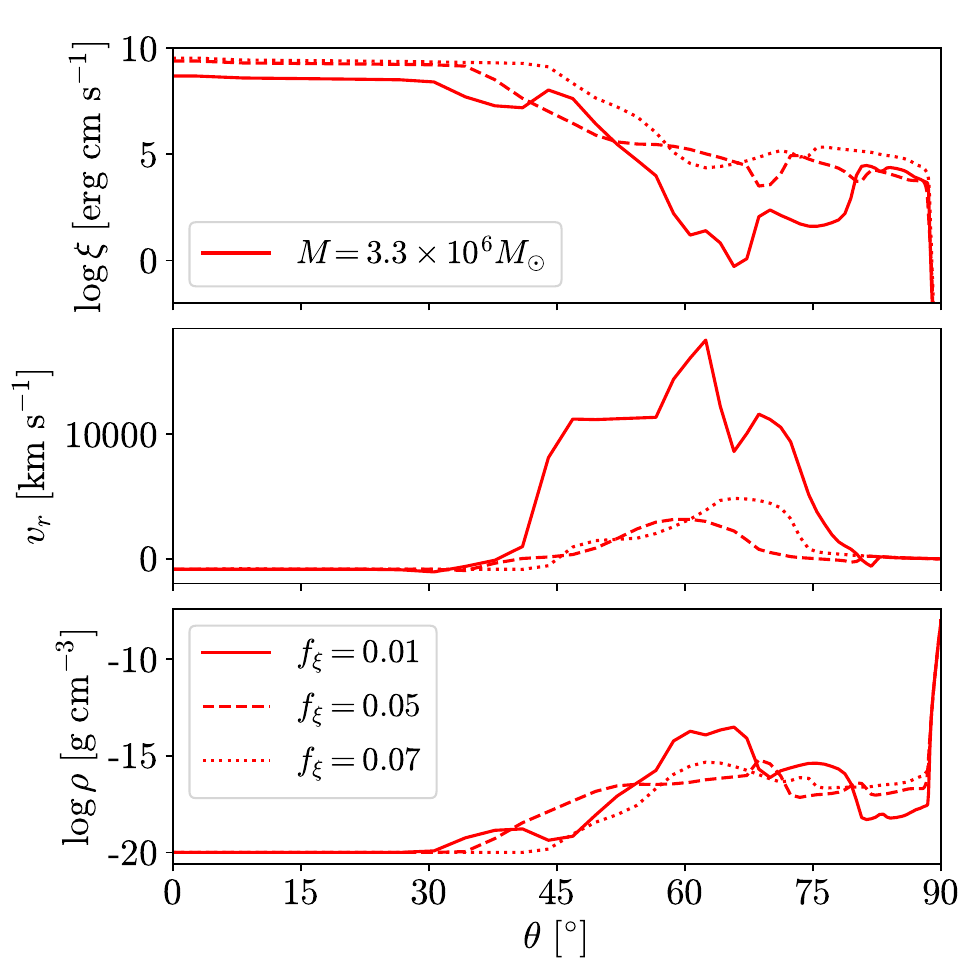}
    \caption{Ionization parameter (top panel), radial velocity (center panel) and density (bottom panel) as a function of altitude at the outer boundary for models with $M = 3.3 \times 10^6 M_{\odot}$, $\Gamma_{d} = 0.3$ and $f_{\xi} = $ 0.01 (solid), 0.05 (dashed) and 0.07 (dotted). Outflows are stronger only for the weakest ionizations.}
    \label{fig:ionization_boundary}
\end{figure}

We investigate the effect of varying the ionizing luminosity emitted by the central object. In Fig. \ref{fig:mdot_ionization} we plot the wind mass flux as a function of ionizing radiation for black holes with $M/M_{\odot} = 3.3 \times 10^6$ (red) and $10^{8}$ (green) for fixed disc Eddington fraction $\Gamma_{d} = 0.3$. We take this as our fiducial disc luminosity, as we described in the previous section, these models had outflows consistent with the assumption of a steady state disc with fixed accretion rate. 

As in our previous study, we find that the mass fluxes and outflow velocities are similar when scaled to the Eddington rate and inner Keplerian velocity. The lower mass models have slightly stronger outflows, most likely due to the enhanced lin driving flux.

In Fig \ref{fig:ionization_boundary} we plot the ionization parameter, radial velocity and density as a function of azimuthal angle at the outer boundary for $GM = 3.3 \times 10^{6} M_{\odot}$ and $f_{\xi} = 0.01$ (solid), $f_{\xi} = 0.05$ (dashed) and $f_{\xi} = 0.07$ (dotted).  

At moderate ionization levels, $f_{\xi} \lesssim 0.05$, the strength of the outflow is unaffected. However, as $f_{\xi} \lesssim 0.01$, the mass flux increases by an order of magnitude and the outflow velocity nearly triples. The enhanced outflow is due to a force multiplier $M(t) \gtrsim 1$ throughout the entire wind. By contrast, models with higher ionizing flux have winds where the force multiplier drops below unity at certain points in the wind, causing the flux to drop and decreasing the outflow velocity. As $f_{\xi} \gtrsim 0.1$, the outflow weakens, and eventually, the wind fails to launch. This occurs for $f_{\xi} = 0.07$  for $1.0 \times 10^{8} M_{\odot}$ and for $f_{\xi} = 0.1$ for $3.3 \times 10^{6} M_{\odot}$.   

Given our finding that significantly reducing the ionizing radiation could enhance outflows, we tried to drive a wind from discs with $\Gamma_{d} = 0.2$. We reduced $f_{\xi} = 0.01$ followed by $f_{\xi} = 0.001$. These conditions were nonetheless unable to drive a wind. 

These results are consistent with previous findings of line-driven disc winds. When the radiation force is barely able to overcome the force of gravity, i.e., when $\Gamma_{d} M_{max} \lesssim 1$, where $M_{max}$ is the maximum value the force multiplier can attain, the wind strength as measured by the mass flux and velocity tends to drop precipitously near this limit. Our current prescription for the line force introduces an ionization dependence on the wind, but this does not change the basic physical picture that when this limit is approached, either due to a weakening of the overall number of line driving photons or the force multiplier due to ionization effects, the wind will fail.

\subsection{SED Effects}
\label{sec:SEDEffects}

\begin{figure}
    \centering
    \includegraphics[scale=0.42]{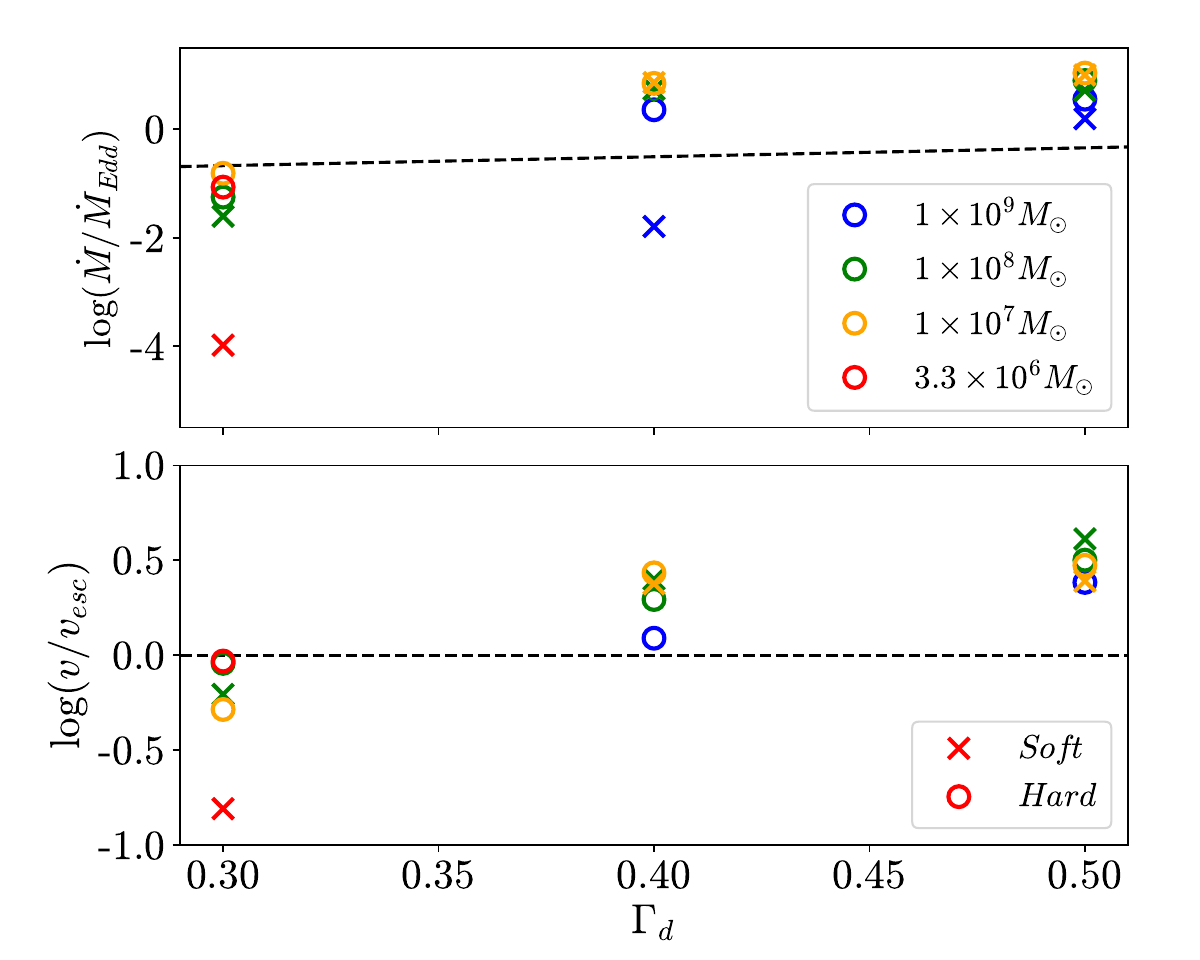}
    \caption{\textit{Top panel -} Wind mass flux in units of Eddington rate (top panel) and terminal outflow velocity (bottom panel) as a function of disc Eddington parameter for black hole masses $M/M_{\odot} = $ $3.3 \times 10^6$ (red), $1. \times 10^7$ (orange), $1. \times 10^8$ (green) and $1. \times 10^9$ (blue). We show results for models with the hard SED  (circles) and soft SED (crosses). The dashed line shows the assumed disc accretion rate.}
    \label{fig:SED_effects}
\end{figure}

\begin{figure}
    \centering
    \includegraphics[scale=0.5]{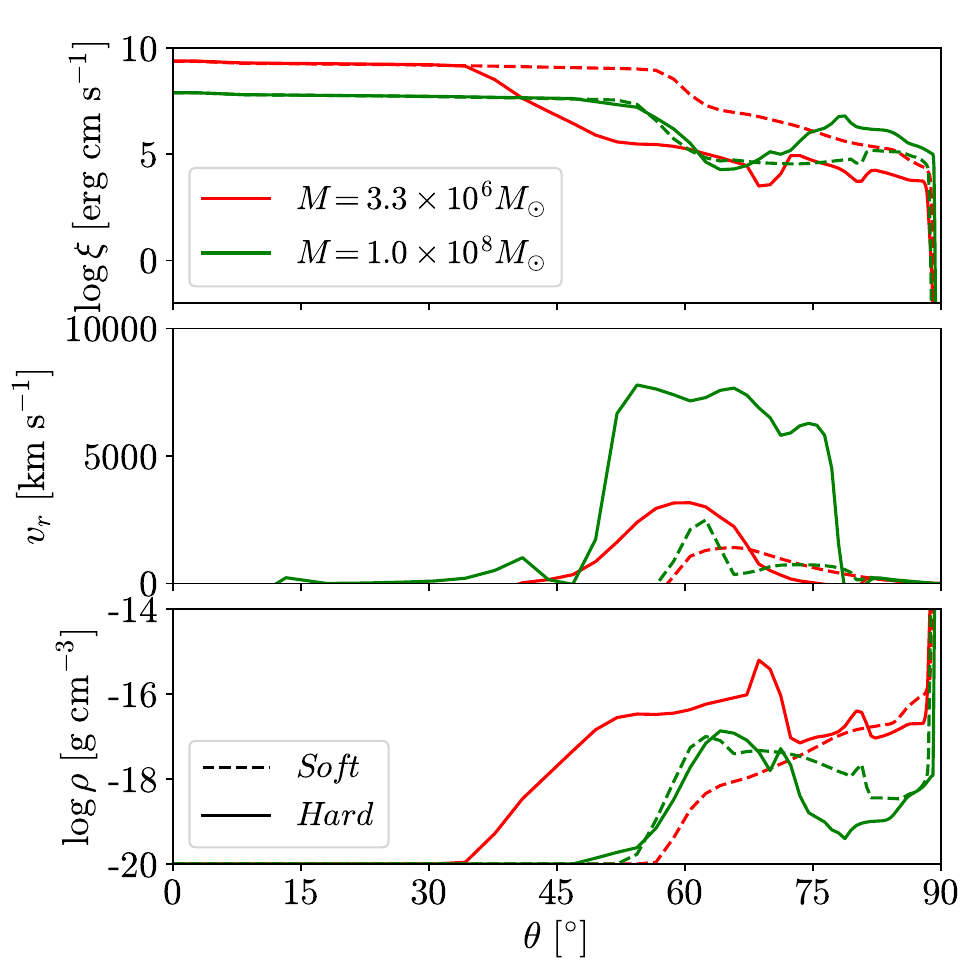}
    \caption{Ionization parameter (top panel), radial velocity (middle panel) and density (bottom panel) as a function of altitude at the outer boundary for models with $M = 3.3 \times 10^6 M_{\odot}$ (red lies) and $M = 1.0 \times 10^8 M_{\odot}$ (green lines) for $\Gamma_{d} = 0.3$ and $f_{\xi} = $ 0.05 for hard (solid lines) and soft (dashed lines) SEDs.}
    \label{fig:SED_boundary}
\end{figure}

We studied the effect of varying the wavelength cutoff for the line-driving band of the radiation field. As explained 
in Sec.~\ref{sec:photoionization}, the latest photoionization studies suggest that photons in the X-ray band can also contribute to line driving. Hence, to better understand the potential implications of this band to AGN winds, we compare models with different lower wavelength cutoffs for the line-driving photons. We consider a SED where the lower frequency cutoff is $\lambda_1 = 200$ \AA, which we will refer to as the {\it soft} SED. This corresponds to the line driving treatment of PK04 and Paper I. We compare these models to those with cutoff $\lambda_1 = 1$ \AA  \, which we will refer to as the \emph{hard} SED. This cutoff is more consistent with our expectations based on our current photoionization modeling. 

In Fig. \ref{fig:SED_effects} we plot the wind mass flux in units of the Eddington rate (top panel) and the outflow velocity in units of the escape velocity (bottom panel) for the soft and hard SEDs for fiducial models with $\Gamma_{d} = 0.3$ and $f_{\xi} = 0.05$ and across the black hole mass spectrum. We find the soft SED models consistently have a lower mass flux than those with the hard SED. This is consistent with our expectation from Fig \ref{fig:rad_field}, which shows the fraction of photons contributing to line driving. This effect is most pronounced for the lowest mass black holes, where $f_{LD} \lesssim 0.5$ interior to the inner radial boundary for the soft SED, whereas the hard SED maintains $f_{LD} \approx 1$ in this region.

We see a similar trend with the outflow velocity, with the lowest black hole mass models with soft SED exhibiting an order of magnitude slower outflow than the hard SED model.  

In Fig \ref{fig:SED_boundary} we plot the ionization parameter (top panel), radial velocity (center panel) and density (bottom panel) as a function of altitude at the outer boundary for models with $M = 3.3 \times 10^6 M_{\odot}$ (red lines) and $M = 1.0 \times 10^8 M_{\odot}$ (green lines) for $\Gamma_{d} = 0.3$ and $f_{\xi} = $ 0.05 for hard (solid lines) and soft (dashed lines) SEDs. We see that the wind driven by the softer SED is strongly suppressed, particularly for the lower-mass black holes.

\section{Discussion}
\label{sec:discussion}

\begin{figure}
    \centering
    \includegraphics[scale=0.5]{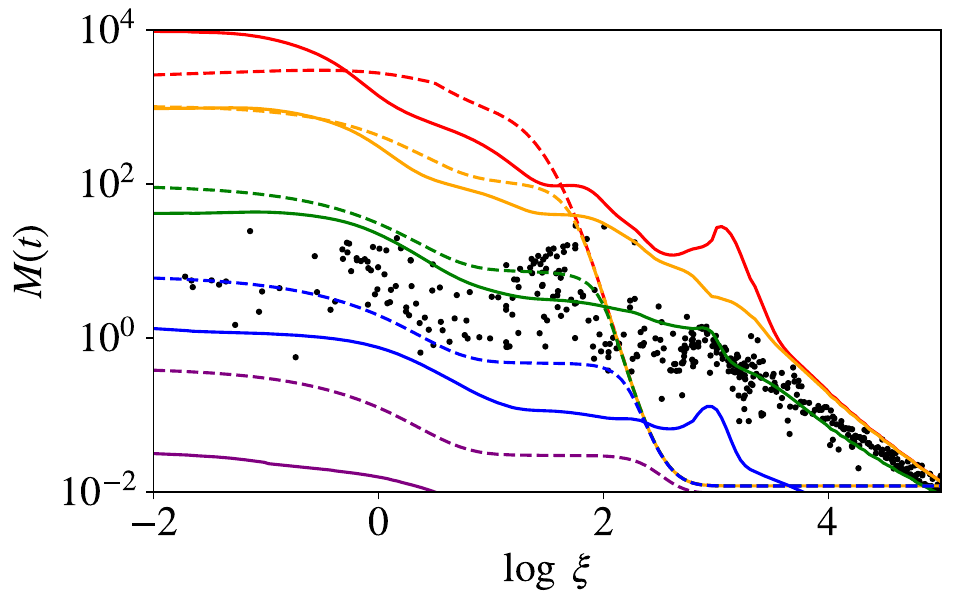}
    \caption{Typical force multiplier phase space diagram for a wind. Our new method for computing force multiplier allows for strong line driving with $\log \xi \approx 3$, where the old analytic model predicted line driving to be weak at these ionizations.}
    \label{fig:phaseSpace}
\end{figure}

We find outflows that are stronger with larger mass fluxes and velocities than in previous studies i.e PSK00, PK04 and Paper I. This is due to an effectively \emph{stronger} line force than used in the above works due to two effects. 

Firstly, as discussed in Sec \ref{sec:photoionization}, the \textsc{XSTAR} photoionization tables have a stronger force multiplier for ionizations $\log \xi \sim 3$ than previously used analytic functions. Fig. \ref{fig:phaseSpace}, shows the typical phase space probed by a wind, where we have overplotted wind data over the force multiplier models shown in Fig. \ref{fig:AGN_Mt}. We note that previous studies of spherically symmetric winds, \citep{Dannen2023}, did not benefit from such a boost, primarily because geometric effects constrain the phase space through which the wind evolves. 

Secondly, the photoionization studies of D19 showed that the number of lines and the distribution of their strengths as measured by the oscillator strength was comparable in the UV and X-ray bands (see their Fig. 2). This finding made us relax the assumption first introduced in \cite{PK04} that only the UV part of the spectrum contributes to the line force. Thus, we moved the lower wavelength cutoff from  $\lambda = 200 \AA$ to $\lambda = 1 \AA$, which we referred to in this work as the soft and hard SEDs, respectively. The new force multiplier tables led to stronger outflows across the entire mass scale. While Paper I found that for $M = 10^{8} M_{\odot}$ and $\Gamma_d = 0.5$ winds were episodic and not particularly strong, this work finds a strong, super-Eddington wind with these parameters. Using a different SED cutoff for the UV radiation had marginal effects on the winds in the center of the mass scale $10^7 \leq M/M_{\odot} \leq 10^8$. However, for $M = 3.3 \times 10^{6} M_{\odot}$ we find that including the high-frequency part of the spectra for the line force allows for winds as strong as for other black hole masses when scaled to the Eddington parameter. These results can be understood qualitatively from the plots of UV fraction Fig \ref{fig:rad_field}. In the inner disc, which is primarily responsible for driving the outflow, black holes on the lower end of the mass range have discs with emission dominated by wavelengths $\lambda \lesssim 200$ \AA.

Models with low ionizing flux fraction $f_{\xi}$ fall into the regime where the force multiplier is approximately constant as a function of ionization, i.e., the top left corner of the left panel of Fig \ref{fig:AGN_Mt}. The force multiplier is approximately saturated to a maximum value, often referred to as $M_{\rm{max}}$ in the line-driven wind literature (see \cite{Owocki1988}). In this case, wind launching will only be possible for disc Eddington fractions $\Gamma_d \gtrsim 1/M_{\rm{max}}$. This explains why even with the lowest values of $f_{\xi}$, we do not see winds with $\Gamma_d \lesssim 0.2$. Increasing the ionization fraction has minor effects on the outflow until the flow reaches $\log \xi \sim 3$, where the force multiplier varies strongly with ionization.  

Many of the cases explored in this study had mass outflow rates higher than the assumed accretion rate of the disc. The assumption that the disc could thus provide a matter reservoir and also irradiate the flow to drive the outflow is inconsistent. Previous studies have tried incorporating some feedback effect of the outflow on the accretion disc. \cite{Nomura2020} used an iterative procedure whereby they let the wind develop, measured the mass outflow rate through the boundary and then changed the disc accretion rate until a self-consistent solution was found. This procedure can work to find stationary solutions, but as we have shown, line-driven AGN winds are highly time-dependent. Further, to leverage future time-dependent observations of AGN from, for instance, the Rubin observatory, we can benefit from such time-dependent simulations. An alternative approach is to modify the disc structure as mass is launched away from the disc. Such an approach was used in \cite{Kirilov2023}, whereby any mass loss from the disc was subtracted from the assumed local disc accretion rate and the intensity was effectively reduced. They applied this approach to low $\Gamma_d \lesssim 0.1$ sources. In the high Eddington regime, such an approach is insufficient, as the mass loss is too high and will affect the global disc structure, not just the local region where the matter has launched. We plan to implement such an approach, since recalculating the global disc structure is the most physically reasonable approach until we have the computational resources to also simulate the full accretion disc.

Including the radiation transfer of the X-rays leads to the development of a complex ionization structure of the gas. Compare, for instance, the right panel in Fig \ref{fig:2dDensityXipanel} to the ionization structure of Fig 1 in PK04. Improvements in computing the force multiplier must likewise follow improvements in computing the ionization parameter, in this case through the radiation transfer of the X-rays.

Despite our efforts to use the most sophisticated photoionization modeling, our approach can be improved in a number of ways. Our current approach accounts for a loss in radiation driving intensity due to geometric effects, but our prescription for the force multiplier assumes a \emph{global} SED, irradiating an optically thin wind. \cite{Higginbottom2024} showed using Monte Carlo radiation transfer simulations, in the context of accreting white dwarf systems, the strength of the force multiplier can decrease due to a changing SED. A key conclusion of their work is that the photons responsible for line driving also contribute to overionizing the gas to the point where the force multiplier is lowered. Their results suggest that such effects may not be sufficiently captured using a single ionization parameter, as the radiation force due to lines is sensitive to the occupation levels of the relevant ions, which in turn depend on the shape of the irradiating SED. They speculate that such effects will be further exacerbated in AGN winds where the structure of the radiation field is more complex. Our model uses a simplified geometric setup where line-driving photons are only emitted from the disc, and ionizing photons are emitted from the central source. A possible future approach is to account for the geometric effects of the irradiating SEDs on the force multiplier. Smith et al. 2023 developed a rapid method for computing the position-dependent SED for a wind irradiated by a disc and corona. They found that near the disc, the SED is approximately that of a blackbody at the disc temperature, whereas close to the axis, it is approximately like the global SED used in this work. With these approximations, it becomes computationally feasible to compute heating/cooling and force multiplier tables for position-specific SEDs, i.e., every cell in the simulation domain should use a different force multiplier table. Thus, we can account for geometric effects, albeit in the optically thin approximation.

Alternatively, we can also relax the optically thin approximation for the line driving flux. Using a multiband approach, we can solve the radiation transfer equation for both the radiation field from the central object and from the multi-temperature blackbody disc. Such a method can account for scattering and absorption effects due to the continuum opacities of the line-driving photons. Also, it becomes possible to compute a mean photon energy, thus improving the current ionization parameter prescription that requires an arbitrary cutoff for the ionizing photons. 

In addition, we note that the outflow is quite sensitive to the radiation field in the innermost parts of the disc. This is unsurprising, given the strong $I \sim r^{-3}$ dependence on the disc intensity and the fact that the Planck function peaks at or near these radii. Given this, it is crucial to accurately model the behaviour of the innermost disc components. For instance, do perturbations further out in the disc, say from a failed wind, propagate inwards and if so at what rate? Alternatively, does Compton upscattering/downscattering change the balance of ionizing to line-driving photons?

\section*{Acknowledgments} 
Support for this work was provided by the National Aeronautics and Space Administration under TCAN grant 80NSSC21K0496. The authors acknowledge Research Computing at The University of Virginia for providing computational resources and technical support that have contributed to the results reported within this publication. SD and SWD acknowledge funding from the Virginia Institute for Theoretical Astrophysics (VITA), supported by the College and Graduate School of Arts and Sciences at the University of Virginia.

\section*{Data Availability Statement}
The simulations were performed with the publicly available code \textsc{Athena++} available at https://github.com/PrincetonUniversity/athena
The photoionization calculations were performed with version 2.37 of \textsc{XSTAR} available at https://heasarc.gsfc.nasa.gov/xstar/xstar.html

The authors will provide any additional problem generators and input files upon request.

\bibliographystyle{mnras}
\bibliography{progalab-shared}

\bsp	
\label{lastpage}
\end{document}